\documentclass{elsart3p}
\usepackage{amsmath}
\usepackage{graphicx}

\renewenvironment{thebibliography}[1]{\begin{svbib}{#1}\tolerance=9999}{\end{svbib}}
\makeatletter
\let\oldps@copyright\ps@copyright
\renewcommand\ps@copyright{\oldps@copyright\renewcommand\@oddhead{\hfil\begin{minipage}[t]{3in}\begin{flushright}BNL-76759-2006-JA\\MUC-PUB-338\end{flushright}\end{minipage}}}
\makeatother

\newcommand\Jv{\vec{J}}
\newcommand\zv{\vec{z}}
\newcommand\nuv{\vec{\nu}}
\newcommand\xiv{\vec{\xi}}

\begin{document}
\begin{frontmatter}
  \title{Amplitude Dependence of Time of Flight and its
    Connection to Chromaticity}
  \author{J. Scott Berg\thanksref{doe:bnl}}
  \date{10 July 2006}
  \thanks[doe:bnl]{Work Supported by the United States Department of Energy,
    Contract No.\ DE-AC02-98CH10886.}
  \ead{jsberg@bnl.gov}
  \ead[url]{http://pubweb.bnl.gov/people/jsberg/}
  \address{Brookhaven National Laboratory; Building 901A; P.O. Box 5000;
    Upton, NY  11973-5000}
  \begin{abstract}
    In general, the time of flight of a charged particle in an
    accelerator will depend on the particle's transverse amplitude.
    This effect can become important in machines with large transverse
    emittances, such as muon accelerators.  We will describe the
    effect, its physical origin, and the circumstances where it
    becomes important.  We will then demonstrate that the effect is
    directly related to chromaticity.  We will describe the effect on
    the longitudinal dynamics in various circumstances, including
    linacs and fixed field alternating gradient accelerators (FFAGs).
    We will describe methods for correcting the effect, particularly
    in FFAGs.
  \end{abstract}
  \journal{Nuclear Instruments and Methods A}
\end{frontmatter}
\section{Introduction}
In 2003, Palmer~\cite{muc-279} demonstrated that in a linear
non-scaling FFAG, there was a dependence of the time of flight on the
transverse amplitude.  Since the effect was small compared to both the
RF period and the variation in the time of flight itself, it was hoped
that the effect would not be too significant.  However, at the 2005
FFAG workshop at KEK, Machida~\cite{ffag05:65} demonstrated that
particles with large transverse amplitude were not accelerated in a
linear non-scaling FFAG.  He also showed the same time-of-flight
dependence on the transverse amplitude.

In this paper, we will first demonstrate that the time-of-flight
dependence on transverse amplitude is directly related to the
chromaticity (transverse oscillation frequency dependence on energy)
of the machine.  While this is a fairly straightforward consequence
of Hamiltonian dynamics, it has not generally been of great
interest in accelerator studies because
\begin{itemize}
\item Transverse oscillation amplitudes tend to be small when compared
  to longitudinal oscillation amplitudes.
\item Chromaticity is corrected to near zero in most circular machines.
\item Longitudinal (synchrotron) oscillations will switch the places
  of late and early arriving particles, reducing the impact of the
  late arrival of a particle with large transverse amplitude.
\end{itemize}
This effect becomes interesting for muon machines, in particular,
because
\begin{itemize}
\item Transverse emittances (areas in phase space) are starting
  to become comparable to longitudinal emittances: after cooling
  for a neutrino factory, for instance, they the transverse emittance
  is about 20\% of the longitudinal.
\item Many of the systems, such as the initial accelerating linac
  and the FFAGs for later acceleration, cannot have their
  chromaticity corrected.
\item Those same systems do not have significant synchrotron
  oscillations.
\end{itemize}
Past studies have primarily looked at the effect of the longitudinal
motion on the transverse, in the context of synchro-betatron coupling
and resonances~\cite{sovjetp5:45,cea-54}.

This effect has been studied before, in the context of ionization cooling
studies~\cite{palmer:fnal1296,prstab2:081001,muc-71}.  It manifested itself
in a correlation between the transverse amplitude and the energy, as
will be explained later in this paper.  This paper connects the phenomenon
with chromaticity, and applies the theory to a wider range of systems,
where its effects are even stronger.  In particular, the effects in
non-scaling FFAGs are thoroughly examined.

\section{Physical and Mathematical Description}

\begin{figure*}
  \includegraphics[width=\linewidth]{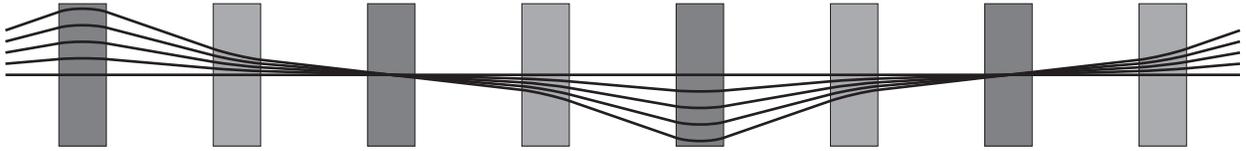}
  \caption{Betatron oscillations with different amplitudes, showing
    which the trajectory length increases with increasing transverse
    amplitude.}
  \label{fig:betaosc}
\end{figure*}%
In its simplest form, when magnets are linear, the time of flight
depends on transverse amplitude because a trajectory oscillating about
the closed orbit has a longer length than the closed orbit itself,
as can be seen in Fig.~\ref{fig:betaosc}.
This arises from the nonzero angles that the oscillations make with
respect to the closed orbit.  For small displacements about the
reference orbit, the length added to the orbit is proportional to the
square of the orbit's displacement.  This makes the time of flight
increase with increasing transverse amplitude.

To compute the effect in general, begin with the Hamiltonian for the
accelerator without RF cavities, $H(\zv,E,s)$, where $\zv$ is a
four-dimensional vector of the transverse phase space coordinates, $E$
is the energy, and $s$ is the arc length along a reference curve that
defines the coordinate system (and is the independent variable).  We
find the energy-dependent closed orbit for the transverse variables
$\zv_0(E,s)$, which is a solution for $\zv$ of Hamilton's equations of
motion and, for a closed ring (or periodic lattice) where the length
of the reference curve for a turn/period is $L$, is a periodic
function of $s$ with period $L$.  We then compute a $4\times4$ matrix
$A(E,s)$ and a function $t_0(E,s)$, which are periodic in $s$, such
that after the transformation from $\zv$ and $t$ (the time) to
$\Bar\zv$ and $\Bar t$ defined by
\begin{equation}
  z = A(E,s)\Bar\zv+\zv_0(E,s)
\end{equation}
\vspace{-\belowdisplayskip}
\vspace{-\abovedisplayskip}
\begin{multline}
  t = \Bar t + t_0(E,s) + \dfrac{1}{2}\Bar\zv^TA^TS(\partial_EA)\Bar\zv\\
  \mbox{}+
  \Bar\zv^TA^TS\partial_E\zv_0+\dfrac{1}{2}\zv_0^TS\partial_E\zv_0,
\end{multline}
the Hamiltonian becomes
\begin{equation}
  \dfrac{2\pi\nuv(E)\cdot\Jv_n}{L}+H_T(E) + O(\Jv_n^{3/2}),
  \label{eq:hnorm}
\end{equation}
where $\Jv_n$ is the two-dimensional vector of action variables,
defined by $J_{n,1}=(\Bar z_1^2+\Bar z_2^2)/2$, $J_{n,2}=(\Bar z_3^2+\Bar
z_4^2)/2$, and $S$ is the 4-dimensional symplectic metric
\begin{equation}
  S = 
  \begin{bmatrix}
    0&1&0&0\\-1&0&0&0\\0&0&0&1\\0&0&-1&0
  \end{bmatrix}.
\end{equation}
I have introduced the notation $\partial_E$, which is the partial
derivative of the following function with respect to $E$.
The subscript $n$ is an indication that I am not using the usual
action variables with dimensions of length, but instead I have not
divided the Hamiltonian and the momenta by the total momentum or a
reference momentum as is usually done, so the dimensions are $\Jv_n$
are energy-time.  $\Jv_n$ should be nearly invariant as one accelerates.
$\nuv$ is the tune as generally defined.

Note that the difference between $t$ and $\Bar t$ is a periodic
function of $s$.  Thus, any time advance from one turn/period to the
next will be governed by the Hamiltonian (\ref{eq:hnorm}).  From
Hamilton's equations of motion
\begin{equation}
  \dfrac{d\Bar t}{ds} = -\partial_EH_T -
  \dfrac{2\pi(\partial_E\nuv)\cdot\Jv_n}{L} + O(\Jv_n^{3/2}).
  \label{eq:dt}
\end{equation}
This is the fundamental theoretical result of this paper: that
the time of flight depends on the transverse amplitude, and the
lowest order dependence is directly related to the tune variation
with energy, also known as the chromaticity.

The result is qualitatively consistent with the original description
of the phenomenon: namely that increased orbit lengths for particles
with large transverse amplitudes increases the time of flight.  Since
for a lattice with no chromaticity correction, the tune decreases with
increasing energy, Eq.~(\ref{eq:dt}) indicates that the time of flight
will increase with increasing transverse amplitude, consistent with
the expected behavior.  Furthermore, Eq.~(\ref{eq:dt}) is linear in
$\Jv_n$, and therefore quadratic in the transverse displacement, again
as expected geometrically.

But interestingly, correction of chromaticity is able to reduce or
eliminate the dependence of time of flight on transverse amplitude.
Sextupoles are introduced into a machine at points with nonzero
dispersion to correct
chromaticity~\cite{cern-isr-as-74-64-42,ieeetns-22-1426,cern-57-21}.
One wants to increase the tune
with increasing energy to correct the chromaticity.  For this
argument, assume a FODO cell with a positive dispersion.  To increase
the horizontal chromaticity, one needs a focusing strength at the
focusing quadrupole (primarily) which increases with momentum.  This
requires a vertical field whose derivative increases with increasing
position, since the dispersion is positive.  This implies a sextupole
which has a vertical field that is parabolic in the midplane and
positive.  Now, consider oscillations in that field.  Since the field
is quadratic, the average field seen by a particle undergoing a
nonzero amplitude horizontal oscillation is positive.  This acts like
an increase in the average dipole field, resulting in an average
radius of the beam which is reduced.  This reduces the length of the
orbit, and thus the time it takes to move along the orbit.  The
average field coming from the sextupole is proportional to the square
of the maximum horizontal displacement in the oscillation.  Thus,
correcting horizontal chromaticity results in a reduced time of flight
for large transverse amplitudes due to the chromatic correction
sextupoles, and consistent with Eq.~(\ref{eq:dt}).

For vertical chromaticity correction, one should have a sextupole at
the defocusing quadrupole (primarily) which has a horizontal field
whose vertical gradient decreases with increasing horizontal position.
From Maxwell's equations, this means that the second derivative of the
vertical field with respect to vertical position must be positive.  As in the
horizontal case, vertical oscillations will thus increase the average
bending field proportionally to the square of the vertical
displacement, resulting in a lower average beam radius and a lower
time of flight.

This result is a useful theoretical tool for two reasons.  First of
all, in the early design stages of accelerator subsystems, one
generally computes energy-dependent closed orbits and the tune as a
function of energy.  Since that information is available already, it
allows one to immediately estimate the effect on the time of flight.
Computing it directly requires tracking for a good distance at nonzero
transverse amplitudes, and averaging out oscillatory effects.  This
requires some trial and error to insure that one is getting the
desired result.  In any case, it is an extra step, when one could have
computed the result without performing it.  Secondly, one immediately
knows one way to go about correcting the effect: by correcting the
chromaticity.  The procedures for correcting chromaticity are
well-known and straightforward.

\section{Applications to Accelerator Systems}

We now examine the effect of the amplitude dependence of the time
of flight on various accelerator subsystems, all of which appear
in most muon acceleration systems.  We will consider motion in a
stationary RF bucket, acceleration in a linac, and acceleration in
a non-scaling FFAG.  We will obtain some rough numerical results
for muon systems with reasonable parameters to give an idea
of the significance of the effect.

\subsection{Stationary RF Bucket}

For a stationary RF bucket, one first wants to find the fixed point
about which one oscillates.  This can be done using Eq.~(\ref{eq:dt}),
by finding the solution where $d\Bar t/ds=h/(f_{\text{RF}}L)$,
where $f_{\text{RF}}$ is the RF frequency, and $h$ is the harmonic
number.  Say that $E_0$ is the
energy where $\partial_E H_T=h/(f_{\text{RF}}L)$.
Expanding $\partial_EH_T$ to first
order near $E_0$, one finds that for small $\Jv_n$, the fixed point
energy is
\begin{equation}
  E_0 - \dfrac{2\pi(\partial_E\nuv)\cdot\Jv_n}{L\partial_E^2H_T},
\end{equation}
where all the derivatives are evaluated at $E_0$.  In terms of the usual
quantities, this is
\begin{equation}
  E_0 + \dfrac{2\pi\xiv\cdot\Jv_n}{T_0\eta},
\end{equation}
where $T_0$ is the time of flight along the closed orbit at energy
$E_0$, $\xiv$ is the chromaticity, defined such that the tune
at energies near $E_0$ is given by $\nuv+\xiv(\Delta p/p)$, with
$p$ being the total momentum, and $\eta$ is the frequency slip
factor.  A similar result holds for a non-stationary bucket.

It seems curious that the effect of the time of flight depending on
amplitude is to have the fixed point energy change with amplitude.
The reason is that the particles always oscillate about a point which
is synchronized with the RF frequency.  Thus, the fixed point comes
about by adjusting the energy so as to adjust the time of flight for
particles with a finite transverse amplitude so that they are
synchronized with the RF.  Therefore, for a group of particles
arriving at the same time and energy at the zero-amplitude fixed
point, the particles with large transverse amplitudes will start to
oscillate about the large-amplitude fixed point, effectively
increasing their longitudinal emittance.
The entire RF bucket will shift as well because of this effect.
Precisely how will depend on how $H_T$ and $\nuv$ vary with energy.

Let's take the example of a muon cooling
lattice~\cite{prstab9:011001}.  The chromaticity is about 0.24, and
$\eta=-1/\gamma^2$, with $\gamma$ being the
ratio of the total energy to the muon's rest mass energy.  The cell
length is about 75~cm.  The maximum normalized transverse amplitude at
the beginning of the channel is about 60~mm (which is $2\Jv_n/mc$
where $m$ is the muon mass and $c$ is the speed of light).  The
resulting closed orbit shift at that amplitude and a reference momentum
of 220~MeV/$c$ is about 31~MeV.  Considering that the full energy
spread of the beam extracted from the cooling channel is about
$\pm 46\text{ MeV}/c$, this is significant.
Those who have studied ionization cooling have been aware of this
effect~\cite{palmer:fnal1296,prstab2:081001,muc-71}, and have found
it important to correctly place an amplitude-energy correlation
into a beam before it enters a cooling channel.

\subsection{Linac}

Consider a linac where the particles are relativistic enough that
synchrotron oscillations can be neglected.  This is true for muons
once they have reached a few hundred MeV.  Say that the tune per cell
in a linac is adjusted to be independent of the momentum as the beam
is accelerated (if it is not, see the next subsection).  If one assumes a
constant accelerating gradient, then a particle with nonzero
transverse amplitude arrives at a time differing from that of a zero
amplitude particle by
\begin{equation}
  -\dfrac{2\pi}{\Delta E}\ln\left(\dfrac{p_f}{p_i}\right)\xiv\cdot\Jv_n,
  \label{eq:linac}
\end{equation}
where $p_i$ is the initial momentum and $p_f$ is the final momentum
(for the entire length of linac), and $\Delta E$ is the energy gain
per cell, assuming that $\xiv$ is the chromaticity for that same cell.

Taking, for example, the last section of the low energy acceleration
linac from~\cite{prstab9:011001}, which accelerates muons from around
520~MeV to 1500~MeV, considering particles with a normalized
transverse amplitude of 30~mm, and assuming a chromaticity of -0.25,
the large transverse amplitude particles are behind the
zero transverse amplitude particles by 0.4~ns.  For 201.25~MHz RF, this
is about $30^\circ$ of phase.  This is comparable to the bunch
length that we are accelerating.  Thus, one should take
these high amplitude particles into account when designing the linac,
to try to ensure that their energy gain is not so different from the
low amplitude particles.  This will be complicated, however, by the
desire to ensure that the all the particles in the longitudinal
distribution also gain roughly the same energy.

Increasing the energy gain per cell reduces the time of flight variation
with transverse amplitude, essentially because the number of cells
the particles traverse is inversely proportional to the energy gain
per cell, and the time of flight variation is essentially proportional
to the number of cells traversed.

\subsection{FFAGs}

FFAGs behave in many ways like a linac, but there are some important
differences.  First, because an FFAG is a multiturn machine, the tune
of each cell is not in general (except for scaling FFAGs) adjusted to
be the same, so Eq.~(\ref{eq:linac}) is not quite correct.  Secondly,
the RF dynamics can be more
complicated~\cite{snowmass2001:T503,epac02:1124,pac03:1831,prstab9:034001}.

The time of flight difference for large amplitude particles can be
computed, assuming a uniform accelerating gradient, to be
\begin{equation}
  -\dfrac{2\pi(\Delta\nuv)\cdot\Jv_n}{\Delta E},
  \label{eq:ffag}
\end{equation}
where $\Delta E$ is the energy gain per cell, and $\Delta\nu$ is the
change in the tune per cell from the initial to the final energy.
Just as in a linac, the effect is reduced if the amount of acceleration
per cell is increased.  The effect can also be improved by reducing
the range of tunes over the acceleration range.  One in principle has
much greater control over the tunes in an FFAG than in a linac, since there
is bending and therefore the opportunity to correct chromaticity.
In fact, the scaling FFAG~\cite{pr103:1837} allows one to in principle
eliminate the tune variation with energy entirely, eliminating
the time of flight variation with transverse amplitude to lowest
order.  For various reasons not discussed here, however, there
are reasons to consider FFAGs where the tune does vary with energy,
in particular linear non-scaling FFAGs \cite{mumu4:693,pac99:3068}.

Let's consider the example of a 10--20~GeV linear non-scaling FFAG for
muons, with a transverse normalized amplitude of 30~mm, an energy
gain of 12.75~MeV per cell, and a tune range of 0.21.  This results a
0.55~ns time slip over the acceleration cycle for a high amplitude
particle when compared to a zero amplitude one.  For 201.25~MHz RF,
this corresponds to a 40$^\circ$ phase slip.

In addition, FFAGs tend to use multiple stages to perform their
acceleration.  Thus, a second stage will tend to increase the time
difference even more.  If one could perform half a synchrotron
oscillation between stages, then in principle the low and high
amplitude particles would switch places and would come back together
at the end of two stages.  However, constructing such a synchrotron
oscillation would require an additional ring (or arc) with RF
voltage comparable to that required for a single FFAG stage, and thus
becomes extremely costly.

One could attempt to correct the effect by creating a positive
chromaticity in the transfer line between the FFAG stages.  One can
compute the required chromaticity for the transfer line using
Eqs.~(\ref{eq:dt}) and (\ref{eq:ffag}):
\begin{equation}
  \xiv = -\dfrac{\beta^2 E}{\Delta E}\Delta\nuv,
\end{equation}
where $\beta$ is the reference velocity divided by the speed of light,
$E$ is the reference energy, $\Delta E$ and $\Delta\nuv$ are as they
were in Eq.~(\ref{eq:ffag}), and $\xiv$ is the chromaticity for the
entire beam line.  The difficulty is the ratio $E/\Delta E$, which is
approximately the number of turns in the FFAG times the number of
cells.  The transfer line cannot expect to create a chromaticity per
cell that is larger than $-\Delta\nuv$, and thus the number of cells
in the transfer line would need to be extremely large for this to
work.

\subsubsection{Chromaticity Correction of the FFAG}
Of course, one could attempt to correct the chromaticity in the FFAG
itself.  However, in doing so, one quickly runs up against what makes
a linear non-scaling FFAG work well: it is able to tolerate the tune
variation despite passing through low-order resonances because
\begin{itemize}
\item the magnets are highly linear, so any resonances are driven
  very weakly,
\item acceleration is rapid, so resonances are passed through very
  quickly, and
\item The lattice is highly symmetric, consisting entirely of short,
  simple, identical cells.
\end{itemize}
Adding nonlinearities has the potential to violate the first of these
conditions.  In fact, any non-scaling lattice with a significant
amount of nonlinearity thus far constructed has failed to have
sufficient dynamic aperture, at least for acceleration of muon beams
that are only modestly cooled~\cite{trbojevic:2003,ffag05fnal:meot,c-a:ap:208}.
Nonetheless, one might hope that the effect can at least be reduced by
adding a modest amount of chromaticity, while simultaneously keeping a
sufficiently large dynamic aperture (with purely linear magnets, the
dynamic aperture is generally much large than necessary, even with the
large dynamic aperture required).

\begin{figure}
  \includegraphics[width=\linewidth]{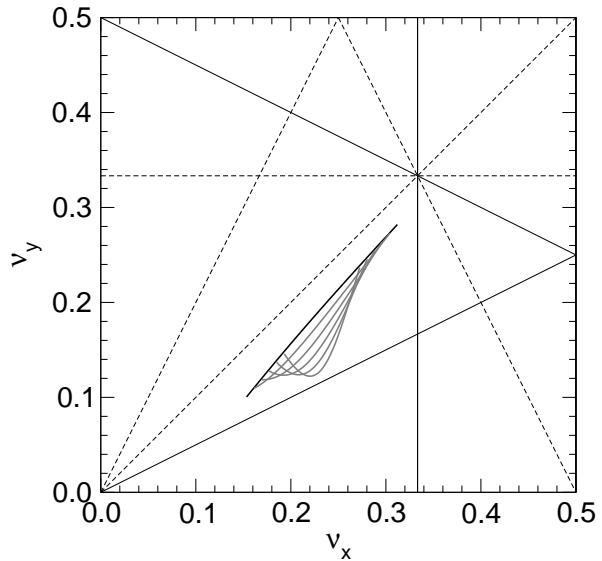}
  \caption{Tune footprint for a non-scaling FFAG for various levels of
    sextupole.  The nearly straight black line gives the tune
    footprint for no sextupole.  The more curved grey lines are for
    increasing levels of sextupole, providing correction from 10\% to
    50\% of the chromaticity.  Other lines indicate resonances up
    through third order, solid lines indicating resonances that are
    driven directly by upright sextupoles, dashed lines indicating
    resonances that aren't.}
  \label{fig:chrom}
\end{figure}
Figure~\ref{fig:chrom} shows the zero-amplitude tune coverage over the
energy range of a non-scaling FFAG for various levels of chromaticity
correction.  A lattice is first constructed which is below the
$3\nu_x=1$ line, and whose tune is equidistant from the
$3\nu_x=1$ and $\nu_x-\nu_y=0$ lines at the low energy end and the
$\nu_x-2\nu_y=0$ line at the high energy end.  This avoids the
resonances that would be directly driven by the sextupole once we add
it, as well as avoiding the linear coupling resonance.  Higher
horizontal tunes are generally preferred for linear non-scaling
lattices~\cite{pac03:2216}, so one is making a sacrifice by staying
below the $3\nu_x=1$ resonance, but adding sextupoles will strongly
drive that resonance~\cite{ffag04triumf:palmer,prstab9:011001}, so it
is necessary.

We then add sextupole and adjust the dipole and quadrupole fields in
the lattice such that the tunes at the low and high energies are given
by
\begin{align}
  \nuv_{\text{lo}}(x) &= (1-x/2)\nuv_{\text{lo}}(0)+(x/2)\nuv_{\text{hi}}(0)\\
  \nuv_{\text{hi}}(x) &= (x/2)\nuv_{\text{lo}}(0)+(1-x/2)\nuv_{\text{hi}}(0),
\end{align}
where $x$ is the fraction by which the chromaticity is corrected.
Figure~\ref{fig:chrom} shows the tune coverage for values of $x$ up to
0.5.  Note that it is really the difference between the low and high
energy tunes that are being corrected, not the chromaticity at a given
energy.  Doing so would have required adding higher order multipoles,
which would have then driven resonances (such as $\nu_x=1/4$) directly
that we have not avoided, leading to potential losses of dynamic
aperture.

Tracking by Machida~\cite{machida:060607} indicates that the dynamic
aperture is acceptable up to $x$ around 0.3.  One does have particle
losses when crossing the $4\nu_x=1$ resonance, but the lattice
parameters can be adjusted to stay just below that.  For higher
amounts of chromaticity correction, the dynamic aperture is reduced at
all energies, so avoiding resonances will not improve the situation.
If a higher level of chromaticity correction were possible, one could
even think about running at a large horizontal tune by staying
entirely above the $3\nu_x=1$ line, but the limited level of
chromaticity correction that seems to be possible precludes that.  One
cannot create multiple sextupole families to try to improve the
dynamic aperture since every cell must be identical (to the extent
possible) to avoid introducing even more resonances.

\subsubsection{FFAG Phase Space Dynamics}

\begin{figure}
  \includegraphics[width=\linewidth]{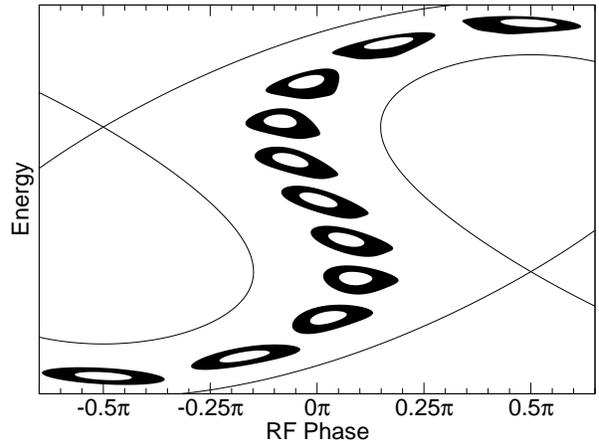}
  \caption{Evolution of a bunch in longitudinal phase space, showing
    the boundaries of the phase space channel.}
  \label{fig:longps}
\end{figure}
\begin{figure}
  \includegraphics[width=\linewidth]{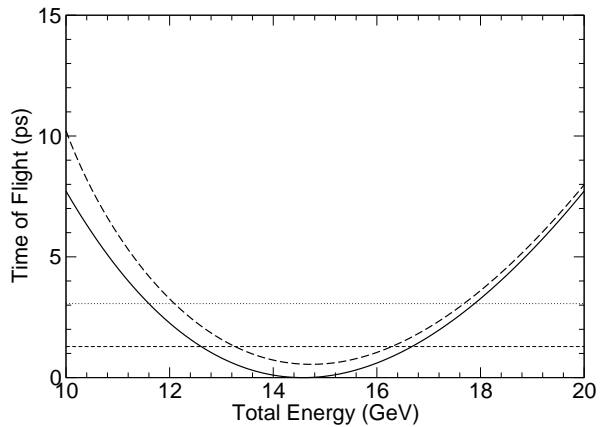}
  \caption{Time of flight as a function of energy for a single linear
    non-scaling FFAG cell.  The solid curve is for zero amplitude, the
    dashed curved is for a horizontal amplitude of 30~mm.  The
    horizontal dashed line is the minimum value of the time that is
    synchronized to the RF for zero amplitude particles, and
    the horizontal dotted line is its maximum value.}
  \label{fig:parab}
\end{figure}
Further complications arise when one considers the longitudinal phase
space dynamics of a linear non-scaling FFAG.  The time of flight in a
linear non-scaling FFAG can be made to be a nearly symmetric
parabola~\cite{pac99:3152} when the particles are highly relativistic.
This leads to particles moving through a serpentine channel through
phase space, as shown in
Fig.~\ref{fig:longps}~\cite{snowmass2001:T503,epac02:1124,pac03:1831,prstab9:034001}.
Figure~\ref{fig:parab} shows what the time of flight as a function of
energy looks like for both zero amplitude particles and particles with
a larger transverse amplitude.  The RF frequency (or the cell length)
is adjusted so that particles with a certain time of flight are
synchronized with the RF.  Particles can only be accelerated over the
desired energy range for a limited range of values for this
synchronization time, which is indicated for the zero transverse
amplitude particles in Fig.~\ref{fig:parab}~\cite{prstab9:034001}.  In
fact, the volume of phase space transmitted is better for certain
values for that synchronization time than others.  Since particles
with a nonzero transverse amplitude have a different time of flight as
a function of energy, the optimal synchronization time will be
different for large amplitude particles than for zero amplitude
particles.  In fact, it may be that there is no value for the
synchronization time for which both low amplitude and high amplitude
particles will be accelerated.

\begin{figure}
  \includegraphics[width=\linewidth]{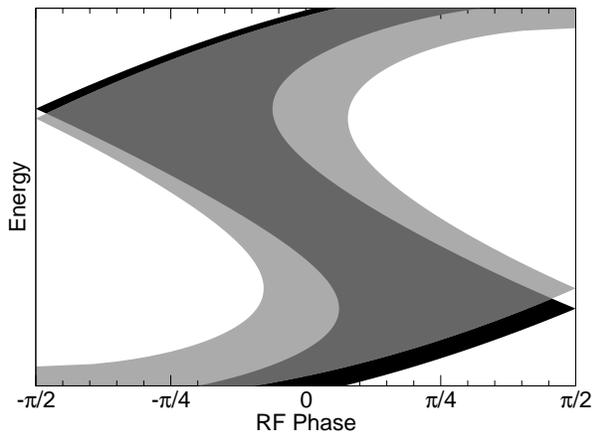}
  \caption{Allowed longitudinal phase space for zero amplitude
    particles (black) and large amplitude particles (light grey).  The
    region where they overlap is indicated in dark grey.}
  \label{fig:longpsamp}
\end{figure}
Another way to look at the problem is to examine the longitudinal
phase space.  Figure~\ref{fig:longpsamp} shows the longitudinal phase
space through which particles of low amplitude will be accelerated, and
the phase space through which particles of high amplitude will be
accelerated, for one sample set of parameters.  Notice the limited
overlap at low energies.  Thus, only a very limited range of phases in
this example can be accelerated, much more limited than the range of
phases one would have expected by only looking at low amplitude
particles.  If the zero amplitude particles alone were used to
determine the initial conditions, many of the high amplitude particles
would not be accelerated at all, which is precisely what is seen
in~\cite{ffag05:65}.  Furthermore, since the initial conditions that
overlap cannot be centered in the channel at either amplitude, there
will be much greater longitudinal distortion of the bunch than one
would have if one were centered in the channel and only had particles
with small transverse amplitudes.  However, this is ameliorated by the
fact that in most cases, one only needs a small longitudinal acceptance
for particles with large transverse amplitudes.

Finally, as discussed before, particles with different amplitudes
clearly arrive at different times, even though they start out at the
same point in longitudinal phase space.  This can be seen from
Fig.~\ref{fig:longpsamp}.  Particles that start out in the overlapping
region at low energy will arrive with a phase close to zero if they
have low transverse amplitudes, but will arrive near a phase of
$\pi/2$ if they had high transverse amplitudes, since the particles
move along lines that are more or less parallel to the left and right
separatrices.  In fact, due to differing times (really numbers of
cells) to traverse the paths, they will also probably arrive with
different energies as well.

\subsubsection{Addressing the Problem in FFAGs}

To address the amplitude dependence of the time of flight in
non-scaling FFAGs, one should probably apply a number of methods, each
of which individually will make the problem somewhat better, and
hopefully together will make the machine perform acceptably:
\begin{itemize}
\item Introduce sextupole components into the magnets to partially
  correct the chromaticity.
\item Create positive chromaticity in the transfer line.  This is less
  likely to give the dynamic aperture problems found in the FFAG since
  one is no longer required to make every cell identical, since one
  only need handle the energy spread of the beam in the transfer line.
\item Introduce second or third harmonic RF into the FFAG.
  This will reduce the
  energy spread in the particles that comes from the different
  longitudinal phase space dynamics that particles with different
  transverse amplitudes have.
\item Increase the amount of voltage per cell.  This helps both
  because the time of flight variation with transverse amplitude is
  reduced (Eq. (\ref{eq:ffag})), but also because the phase space
  channel for the particles gets larger, increasing the overlapping
  area for different transverse amplitudes~\cite{prstab9:034001}.
  Optimized machine designs often left empty cells because doing so
  reduced the magnet aperture and therefore the cost~\cite{ffag04:1}.  It
  appears more important to correct the problem described here.  One
  could even use two-cell cavities rather than single cell to improve
  the behavior further, but this could potentially increase costs
  significantly.
\end{itemize}

\section{Conclusions}

We have demonstrated that the lowest order dependence of the time
of flight on transverse amplitude is directly related to the chromaticity
of a machine.  We have used that fact to quickly estimate its effect
in several accelerator systems.  In particular, we have shown why it is
important in non-scaling FFAGs, and have explored possible methods
of addressing the problem.

\bibliographystyle{elsart-num}
\bibliography{060621}
\end{document}